\documentclass[pdflatex,sn-mathphys-num]{sn-jnl}% Math and Physical Sciences Numbered Reference Style
\newcommand{\dd}{\mathrm{d}}
\newcommand{\pd}{\partial}

\newcommand{\ub}{\boldsymbol{u}}

\newcommand{\degree}{^\circ}

\usepackage{graphicx}%
\usepackage{multirow}%
\usepackage{amsmath,amssymb,amsfonts}%
\usepackage{amsthm}%
\usepackage{mathrsfs}%
\usepackage[title]{appendix}%
\usepackage{xcolor}%
\usepackage{textcomp}%
\usepackage{manyfoot}%
\usepackage{booktabs}%
\usepackage{algorithm}%
\usepackage{algorithmicx}%
\usepackage{algpseudocode}%
\usepackage{listings}%
\usepackage{array}
\usepackage{caption, subcaption}
\usepackage{euscript}

\theoremstyle{thmstyleone}%
\theoremstyle{thmstyletwo}%

\theoremstyle{thmstylethree}%

\begin{document}

\title[Article Title]{Lift and leading-edge suction parameter of separated flows over an NACA0012 at high angles of attack}


\author*[]{\fnm{Ching} \sur{Chang}*}	\email{c.chang@pme.nthu.edu.tw}

\author{\fnm{You-Peng} \sur{Shih}}	%\email{iiauthor@gmail.com}

\author{\fnm{Tang-An} \sur{Li}}		%\email{iiiauthor@gmail.com}

\affil{\orgdiv{Power Mechanical Engineering}, \orgname{National Tsing Hua University}, \orgaddress{\street{No. 101, Section 2, Kuang-Fu Road}, \city{Hsinchu}, \postcode{300044}, \state{}, \country{Taiwan}}}


\abstract{The flow condition at the leading edge governs the dynamics of the leading-edge vortex, which is crucial for understanding the separated flow over an airfoil at high angle of attack. Furthermore, with extensive applications in biomimetic flight, the wings encountering high-angle-of-attack situations in an unsteady manner are of great interest. The leading-edge suction parameter (LESP) is a dimensionless metric proposed to quantify the leading-edge flow condition, and is implemented in the LESP-modulated discrete vortex method, which successfully predicts aerodynamics of airfoils in motion. To discern the timing of leading-edge vortex formation, a critical threshold for LESP is chosen to control the onset of separation. However, it is not obvious that the same strategy could be applied to a stationary wing where the separation is not dominated by the motion of the airfoil. We conduct computational fluid dynamics (CFD) simulations for a stationary NACA0012 airfoil at high angles of attack, and extract the leading-edge flow quantities from the CFD data. In addition, vorticity flux, which contributes to the formation of vortices above the top surface of the airfoil, is also investigated to reveal the vorticity budget and its relevance to aerodynamic performance. We show that for the laminar case ($Re=1000$), the instantaneous LESP is well correlated with the lift, while for the turbulence ($Re=10^5$), the time-averaged LESP is well correlated with the lift. The result would provide insights into future improvements for vortex-based models of separated flows.}

\keywords{NACA0012, high angle-of-attack, leading edge, separated flow}



\maketitle

\section{Introduction}\label{sec1}
The separation flows and the vortices generated when a fluid flows past a slender body at a high angle of attack constitute a significant topic in fluid dynamics \cite{Deparday2022, Khalifa2023, Chen2025}. Among these, the flow around airfoils is particularly complex due to the involvement of various aerodynamic configurations. Leading-edge vortices (LEVs) \cite{Eldredge2019} typically occur during high-angle-of-attack attitude. The boundary layer develops near the leading edge, then it separates due to an adverse pressure gradient and becomes a free shear layer \cite{Deparday2019}. The free shear layer curls up to form an LEV. Depending on the Reynolds number, this can be stable or unstable. Such vortices frequently appear in low-Reynolds-number \cite{Durante2020, Chen2025}, unsteady aerodynamics, such as the biological flight of insects and birds. Trailing-edge vortices (TEVs) are vortex structures generated when fluid passes over the airfoil trailing edge due to pressure differences and viscous effects. Like leading-edge vortices, they are classified as stable or unstable vortices based on the Reynolds number. A trailing-edge vortex is a vortex structure generated by pressure differences and viscous effects as fluid passes over the wing's trailing edge. Like the leading-edge vortex, it represents separated flow and becomes more pronounced at high angles of attack or during stall condition. Secondary vortices refer to vortices that form near the primary vortex (such as the leading-edge vortex) due to the swirling motion of the main vortex. The fluid interacts again with the wall surface, forming a boundary layer in the opposite direction. These vortices are also induced by pressure differences and viscosity. Their direction is opposite to that of the primary vortex and they typically appear on the wing surface and below the primary vortex. As the Reynolds number and angle of attack increase, the flow field evolves from a steady state to periodic vortex shedding, ultimately exhibiting irregular and chaotic characteristics. The generation, development, and interaction of various vortices become key research focuses \cite{Akkala2017}, as they not only affect flow stability but also significantly influence aerodynamic performance \cite{Li2023}. 

Leading-edge suction parameter (LESP) is first proposed by the development of discrete vortex method (DVM) \cite{Ramesh2014}. The method is based on the classic thin airfoil theory \cite{Katz2001}, where the airfoil is modeled by a two-dimensional vortex sheet. The strength of the vortex sheet is forced to zero at the trailing edge in order to satisfy the Kutta condition; while the leading edge can support a non-zero circulation. The leading-edge suction parameter, physically the non-dimensional suction force, is proposed to modulate the release of vorticity from the leading edge. The LESP-modulated discrete vortex method (LDVM) has been ever since implement as a reduced-order model to study unsteady aerodynamics \cite{Ramesh2014, Ramesh2018, Narsipur2020}. It has been shown that for high pitching rate ($K>0.1$), the critical LESP, at which the LEV initiates, is approximately constant at a given Reynolds number and airfoil shape. However, for $K<0.1$ the critical LESP depends on the pitching rate. 

In order to extract the LESP from flow data, a method is proposed to quantify the leading-edge suction \cite{Deparday2022}. By employing a virtual circular contour, the circulation over the leading edge section can be integrated. Subsequently, incorporating the airfoil's curvature radius, chord length, the boundary layer thickness at the intersection point between the contour and airfoil, and the stagnation point location enables the calculation of the instantaneous LESP of the flow field. Two experimental studies were established, one examined perturbed flow over a fixed thin airfoil (e.g., NACA0009). Since the curvature radius at the airfoil leading edge is far smaller than the stagnation point location, the stagnation point effect can be omitted to simplify calculations. Concurrently, they considered the contribution of shear layer thickness on LESP. The shear layer represents the high-vorticity region between the separated flow and the free-stream flow during flow separation. Incorporating its thickness into LESP calculations yields results more consistent with their experimental data. The other experiment involved uniform flow over an oscillating thick airfoil (e.g., NACA0015), where a consideration of the stagnation point location for LESP calculations is necessary. 

A study on the vorticity transport around a heaving flat plate can be found in \cite{Akkala2017}. It combines experimental observations of the flow field around the upper surface of a moving plate with analysis to quantify the contribution of different vorticity fluxes to the total circulation within the control region. Within a single motion cycle, the primary flux sources within the control region are the shear layer flux and the diffusive flux. The former represents the vorticity flux originating from the leading edge, while the latter arises from the boundary layer on the plate surface, involving vorticity generation by viscosity and the adversary pressure gradient.

Extensive studies have been done for maneuvering wing undergoing pitching and/or heaving. The LDVM\cite{Ramesh2014} has thus gained a lot of success in modeling this kind of separated flows. However, it is not obvious that the vortex shedding of a stationary airfoil at high angle of attack can be capture by the LDVM. To investigate this limitation and provide avenues for improvement, this study simulates the flow field of the NACA0012 airfoil under both laminar and turbulent conditions with Reynolds numbers set to $1,000$ and $100,000$. The angles of attack range from $18\degree$ to $30\degree$. By examining vortex structures under different conditions and quantifying the LESP and vorticity flux at the leading edge, this study investigates the relationship between LESP, vorticity flux, and lift in a viscous flow. This serves as a preliminary exploration for applying discrete vortex models to fixed-wing scenarios. A description of numerical treatment is given in \textsection \ref{sec2}, the result and discussion in \textsection \ref{sec3}. Finally, we conclude in \textsection \ref{sec4}.

\section{Numerical methods}\label{sec2}

\subsection{Governing equations}\label{sec2.1}
We numerically solve the unsteady viscous incompressible flows, governed by the Navier--Stokes equation and continuity equation
\begin{equation}
\nabla \cdot \ub =0,
\end{equation}
\begin{equation}
\frac{\pd \ub}{\pd t} + \ub \cdot \nabla \ub = -\frac{1}{\rho}\nabla p +\nu \nabla^2\ub 
\end{equation}
where the density $\rho$ and kinematic viscosity $\nu$ are constant. 

An open-source code OpenFOAM is utilized in the current simulation study. The PISO (Pressure-Implicit with Splitting of Operators) algorithm was selected for calculating the pressure and velocity fields within the flow. The PISO algorithm enables the determination of time-dependent velocity and pressure distributions in the flow field. An initial guess for pressure and velocity values is first set. These initial values are then substituted into the momentum prediction equation to obtain the first velocity field. This velocity field is then used to solve Poisson's equation, which calculates the pressure. This corrected pressure is subsequently applied to the momentum correction equation to obtain an updated velocity field. This result is then fed back into Poisson's equation to correct the pressure again, completing one iteration cycle. Finally, the corrected pressure is applied to the momentum correction equation to obtain the third velocity field result, which is then carried forward to the next computational time step. 

In this simulation study, we investigate both the laminar and turbulent flow regimes. They are characterized by the dimensionless Reynolds number
\begin{equation}
Re = \frac{U_{\infty} c}{\nu},
\end{equation}
where $c$ is the chord length of the airfoil, $U_{\infty}$ is the flow speed of the incoming freestream, and $\nu$ is the fluid viscosity. We set $Re=1,000$ and $100,000$ in our laminar and turbulent simulations, respectively. 

For simulations in the turbulence regime, a turbulence model in the Improved Delayed Detached Eddy Simulation (IDDES) is implemented. DES \cite{Spalart1997, Shur1999} (Detached Eddy Simulation) combines RANS (Reynolds-averaged Navier$-$Stokes equations) and LES (Large Eddy Simulation) models by resolving vortices at different scales to achieve a balance between computational cost and accuracy. It employs LES to simulate large-scale vortices in the separation region away from the wall surface, while utilizing RANS in $k-\omega$ SST or Spalart$-$Allmaras models in regions with higher velocity gradients to better capture wall shear stress and reduce computational cost. DDES \cite{Spalart2006} (Delayed Detached Eddy Simulation) introduces a shielding function to delay the transition from RANS to LES, ensuring wall simulations use RANS rather than LES. IDDES \cite{Gritskevich2012} builds upon this foundation and further enhances the accuracy of model transitions.

 \begin{figure}[h!]
\centering
 \begin{subfigure}[b]{\textwidth}
\centering
\includegraphics[width=0.45\textwidth]{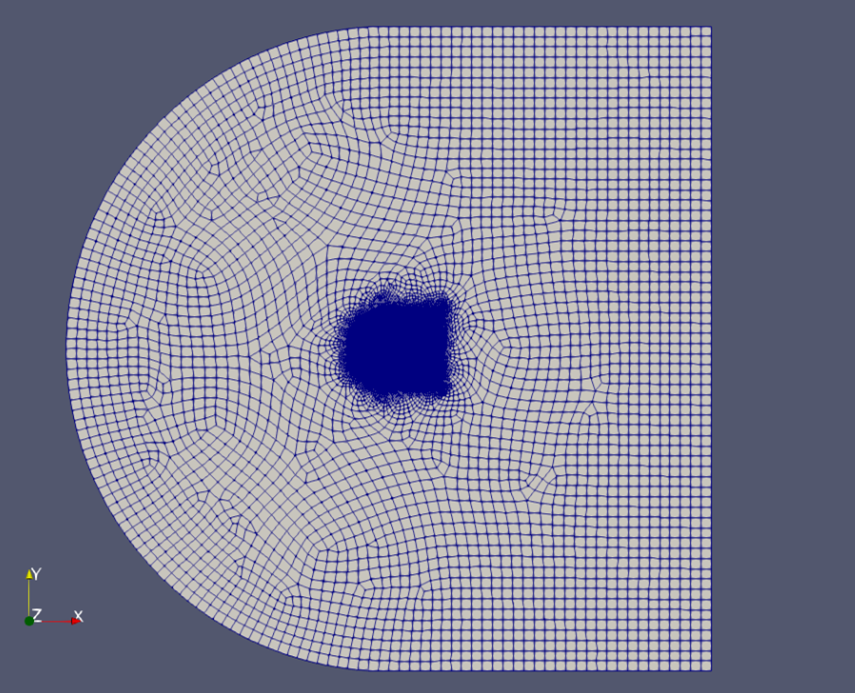}
\includegraphics[width=0.45\textwidth]{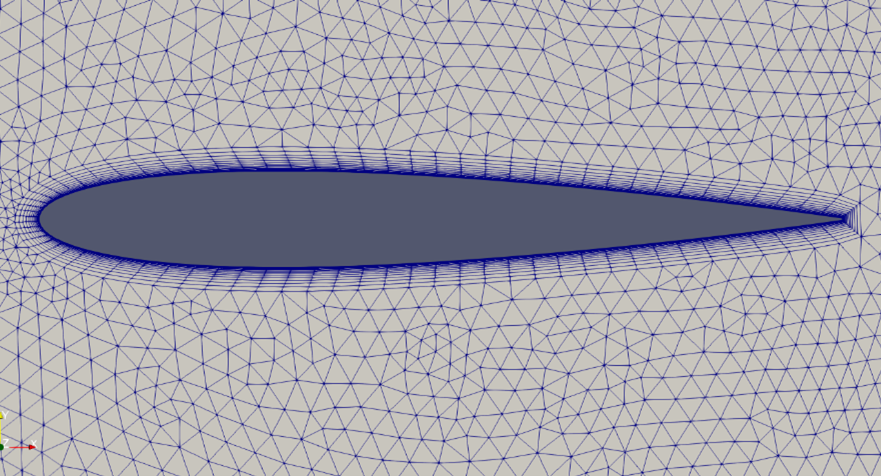}
\caption{}
\end{subfigure}
\\
\begin{subfigure}[b]{\textwidth}
\centering
\includegraphics[width=0.45\textwidth]{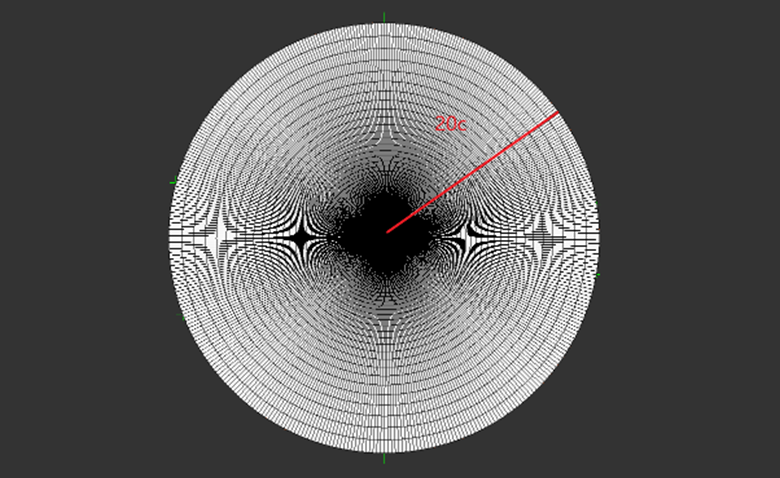}
\includegraphics[width=0.45\textwidth]{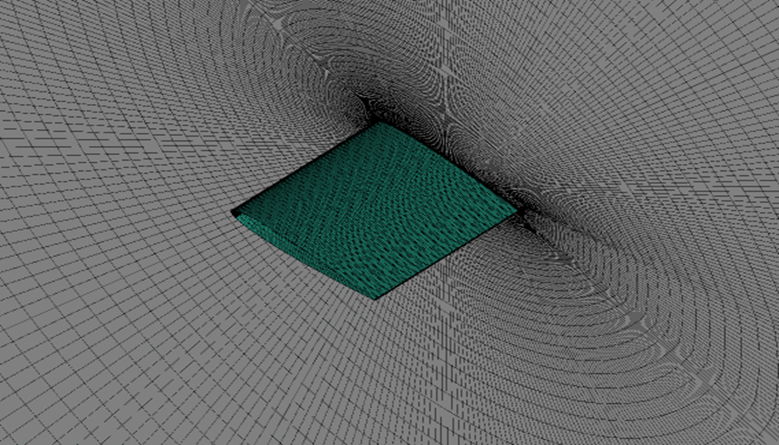}
\caption{}
\end{subfigure}
\caption{The computational mesh for (a) $Re=1000$; (b) $100,000$. Freestream is coming from the left and exists to the right.}\label{fig: domain}
 \end{figure}

\subsection{Computational domain and boundary condition}\label{sec2.2}
We employ a Cartesian coordinate system with the origin positioned at the leading edge of the wing. Inlet conditions are set with the streamwise component of the velocity at $1 ~m/s$ and zero pressure gradient. The exit conditions are set to zero velocity gradient and a constant pressure (gauge $0$). The airfoil surface is no-slip in velocity and has a pressure gradient of zero. The computational domain is a two-dimensional C-shape for the $Re=1000$ laminar simulations, with the upper, lower, left, and right boundaries extending 25 chord lengths from the airfoil, see Figure \ref{fig: domain} a. For turbulence simulations at $Re=10^5$ using IDDES, a three-dimensional O-shape domain \cite{Khalifa2023} of radius $20c$ is used with spanwise length of $1c$ in Figure \ref{fig: domain} b, and periodic boundary conditions are imposed spanwisely.

\subsection{Mesh validation}\label{sec2.3}
\begin{figure}[h!]
\centering
\includegraphics[width=0.7\textwidth]{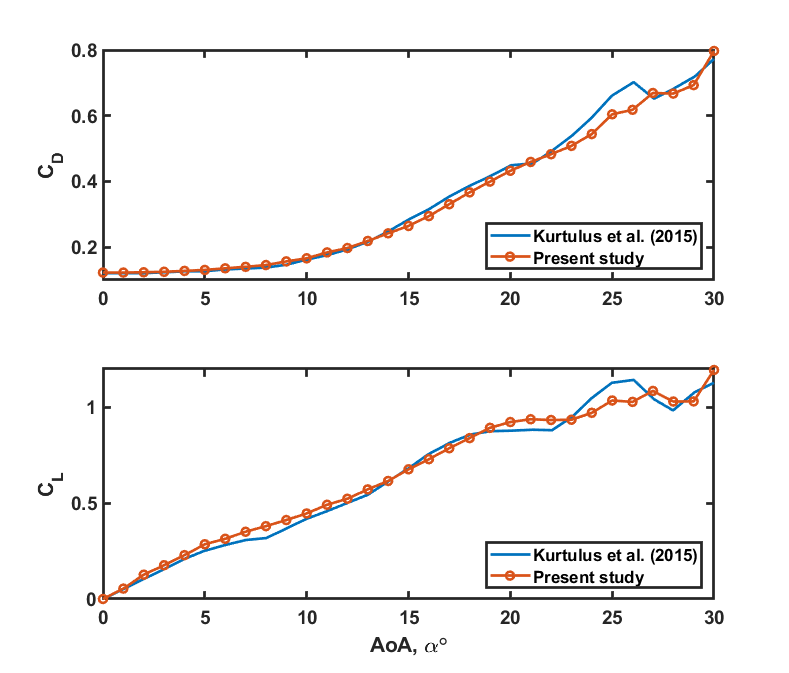}
\caption{Drag and Lift coefficients for $Re=1000$ validated with literature versus angles of attack.}\label{fig: Re1000 valid}
 \end{figure}

Simulation results at Reynolds number $1000$ were compared with literature \cite{Kurtulus2015}, as shown in Figure \ref{fig: Re1000 valid}. The total number of cells is 33,700 for the two-dimensional laminar simulations with $y^+$ being set below $1$. It demonstrates good agreement across different angles of attack. Both coefficients computed from our simulations qualitatively agree with the values reported from the literature at an angle of attack between $0\degree$ to $30\degree$. %The maximum deviation appears near AoA $25\degree \sim 26\degree$, which has error of $10\%$ in the drag and $7\%$ in the lift.

Validation for the simulation at a Reynolds number of $100,000$ is achieved by comparing our result against both the experiment \cite{Ohtake2007} and simulation \cite{Guney2023} over different angles of attack. The number of meshes ranges from 750,000 to 1,000,000, depending on the angle of attack. $y^+$ is controled at approximately $0.6$.

%\begin{table}[h!]
%\caption{Mesh independence study}	\label{tab: mesh indep}
%\begin{tabular}{ p{2cm } p{2cm} p{2cm} l p{1cm} p{1cm} }
%\toprule
%Mesh\newline configurations 	& Cells around\newline the airfoil & First layer\newline thickness (m)	& Total cell number	& Error in $C_D$\footnotemark[1]  &Error in $C_L$\footnotemark[1]\\
%\midrule
%Coarse			& 300				& $1.5\times10^{-4}$	 	& 82389 			&22\% 				&36\%	\\
%Medium 1			& 400			 	& $3.7\times10^{-6}$ 		& 115743 			&12\% 				&8\%		\\
%Medium 2		 	& 500 				& $3.7\times10^{-6}$	 	& 155762			& 6\%				&2\%		\\
%Fine		 		& 800 				& $3.7\times10^{-6}$	 	& 203182			& 4\%				&1\%		\\
%\botrule
%\end{tabular}
%\footnotetext[1]{Compared with the experimental measurement \cite{Ohtake2007} for NACA0012 at AoA $20\degree$.}
%\end{table}

 \begin{figure}[h!]
\centering
\includegraphics[width=0.7\textwidth]{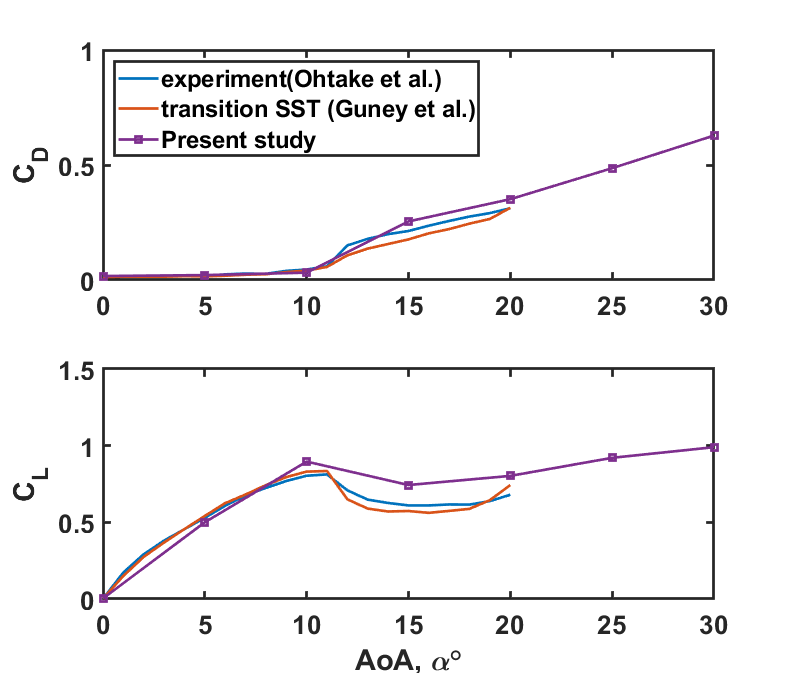}
\caption{Drag and Lift coefficients for $Re=10^5$ validated with experiment and simulation versus angles of attack.}\label{fig: Re1e5 valid}
 \end{figure}

\section{Results and discussion}\label{sec3}
We first analyze the flow dynamic for $Re=1000$ in both time-dependent and time-averaged sense. Then we proceed to extract the leading-edge flow condition in LESP. The analysis for $Re=10^5$ turbulent flow in a time-averaged sense then follows.

\subsection{Laminar regime, $Re=1000$}
\begin{figure}[h!]
\centering
\includegraphics[width=0.7\textwidth]{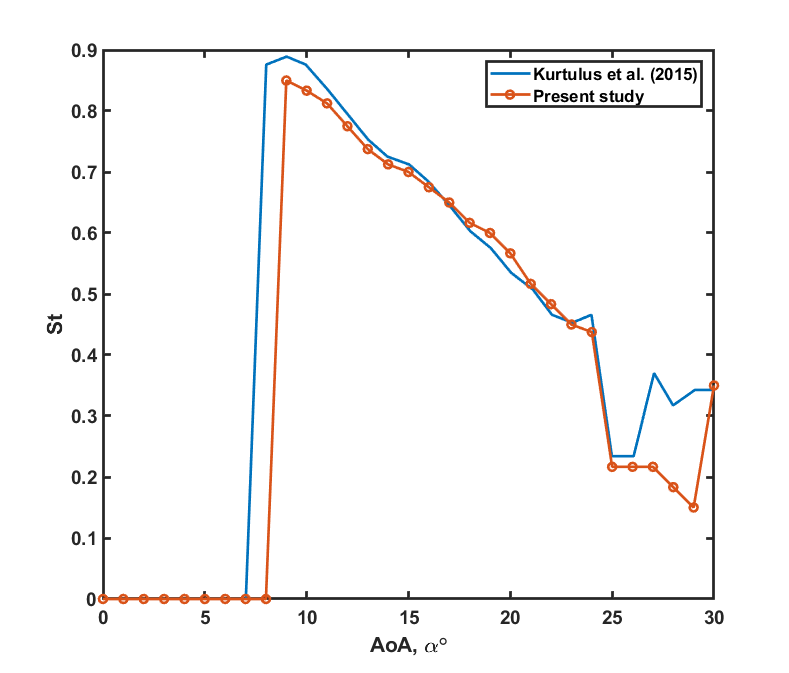}
\caption{Strouhal number at different angle of attack. A comparison with the simulation by \cite{Kurtulus2015}.}\label{fig: Re1000 St}
 \end{figure}

Flow over an NACA0012 airfoil is numerically simulated for angle of attack from $0\degree$ to $30\degree$, the drag and lift coefficients (time-averaged if unsteady flow is present)
\begin{equation}
C_D =\frac{F_x}{0.5\rho U^2_{\infty}c}, ~~~C_L =\frac{F_y}{0.5\rho U^2_{\infty}c}
\end{equation}
are computed and plotted in Figure \ref{fig: Re1000 valid}. Unlike operating at a conventional Reynolds number of $O(10^5\sim10^6)$, there is no significant stall while the angle of attack reaches up to $30\degree$ for $Re=1000$. The vortex shedding kicks in around the angle of attack around $7\degree \sim 8 \degree$, and the Strouhal number
\begin{equation}
St =\frac{f~c}{U_{\infty}}
\end{equation}
is plotted in Figure \ref{fig: Re1000 St}. Comparing our simulation result with literature \cite{Kurtulus2015}, we can find that the onset of vortex shedding differs by only $1\degree$. Once the vortex shedding onset, the trend of Strouhal number up to $24\degree$ agrees well both qualitatively and quantitatively. However, a significant deviation emerges for angle of attack above $26\degree$, while \citep{Kurtulus2015} captures a higher shedding frequency and ours takes a dip. It is noted that the Strouhal numbers match again at AoA $30\degree$.

\begin{figure}[h!]
\centering
 \begin{subfigure}[b]{0.4\textwidth}
\includegraphics[width=\textwidth]{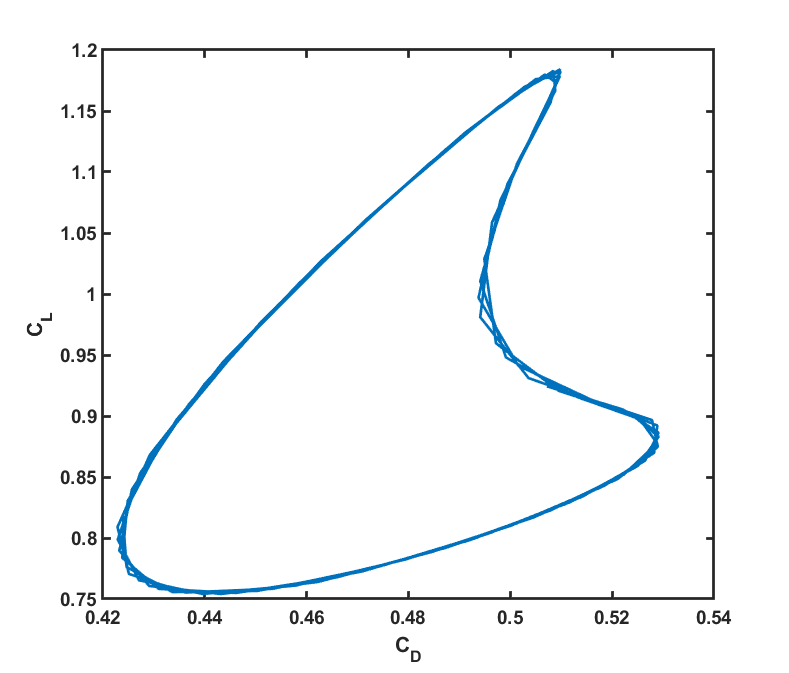}
\caption{AoA $22\degree$}
 \end{subfigure}
~
 \begin{subfigure}[b]{0.4\textwidth}
\includegraphics[width=\textwidth]{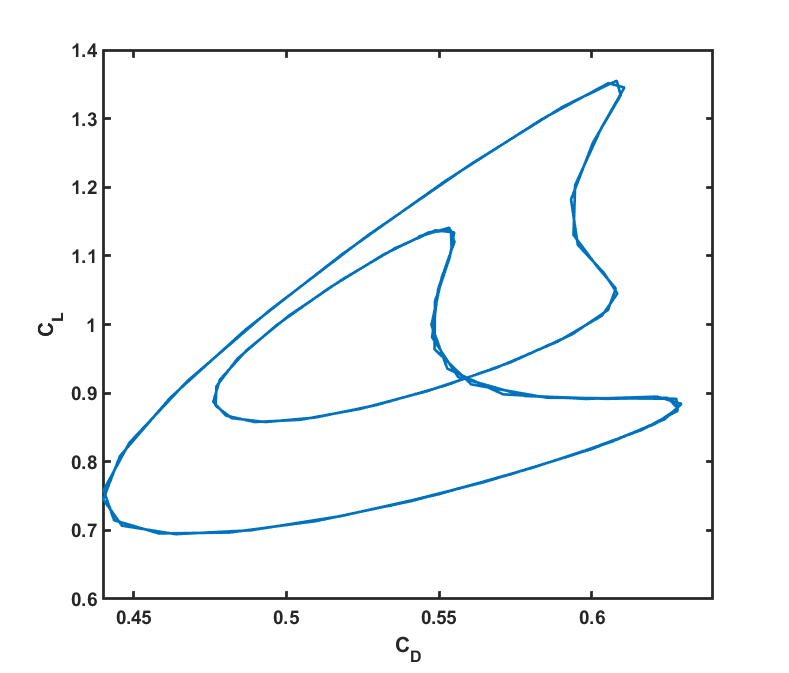}
\caption{AoA $24\degree$}
 \end{subfigure}
~
 \begin{subfigure}[b]{0.4\textwidth}
\includegraphics[width=\textwidth]{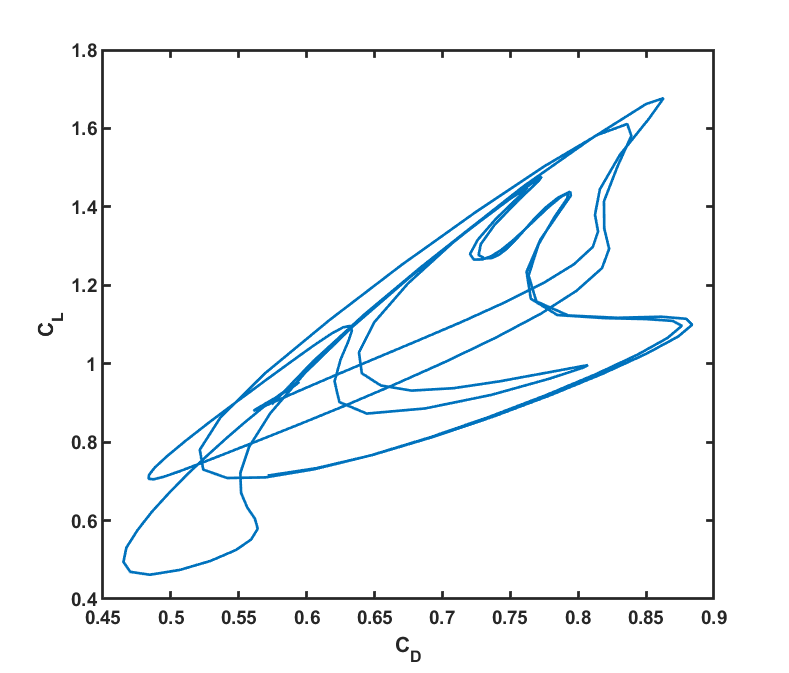}
\caption{AoA $28\degree$}
 \end{subfigure}
~
 \begin{subfigure}[b]{0.4\textwidth}
\includegraphics[width=\textwidth]{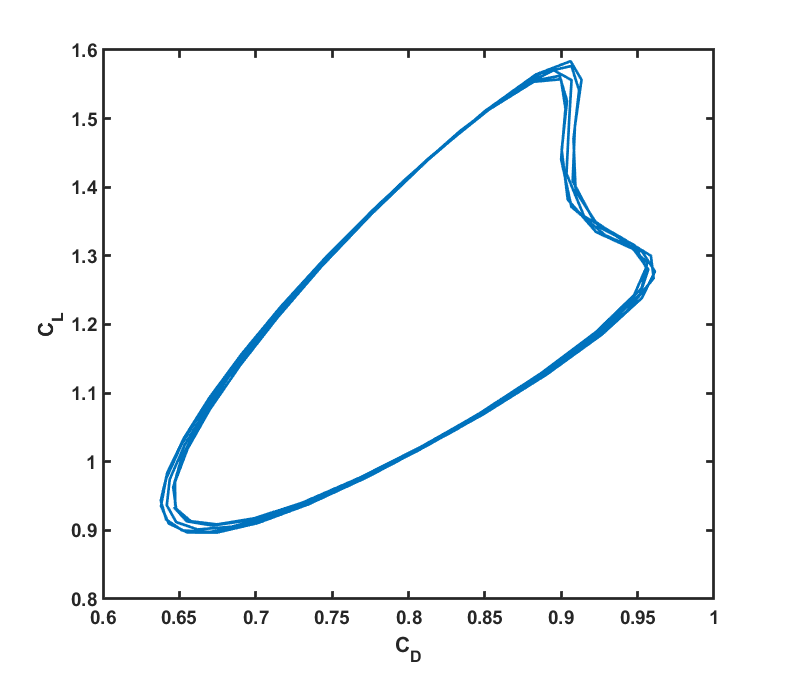}
\caption{AoA $30\degree$}
 \end{subfigure}

\caption{Phase portrait for different angles of attack, representing distinct dynamical regimes.}\label{fig: phase portrait}
 \end{figure}

There is a drastic drop in the Strouhal number for AoA $24\degree \sim 25\degree$, which might indicate the separated flow undergoes a transition to a distinct dynamic. Therefore, we investigate phase portraits by plotting the instantaneous $C_D$, $C_L$, and examining the orbit shown in Figure \ref{fig: phase portrait}. From $22\degree$ to $24\degree$, there is clearly a change in dynamical behavior, where a simple periodic dynamic in $22\degree$ became self-intersecting once. A smaller, secondary period emerges within the larger, primary period. This is referred to as the period-doubling regime. The orbit then turns into unorganized and non-periodic when the angle of attack is $28\degree$. This may explain the deviation in the Strouhal number from our simulation compared to \cite{Kurtulus2015}. The chaotic and non-periodic dynamics might make quantification more challenging, as the dynamics are non-periodic; therefore, computing time-averaged quantities in $C_D$ and $C_L$ within a finite time would not yield a consistent result. Surprisingly, the orbit is back to periodic for $30\degree$ angle of attack, and the predicted Strouhal numbers in Figure \ref{fig: Re1000 St} agree again. The transition from periodic to periodic doubling, then to a non-periodic, chaotic regime, and finally back to periodic, has also been observed in a recent study \cite{Chen2025}.

 \begin{figure}[h!]
\centering
 \begin{subfigure}[b]{0.48\textwidth}
\includegraphics[width=\textwidth]{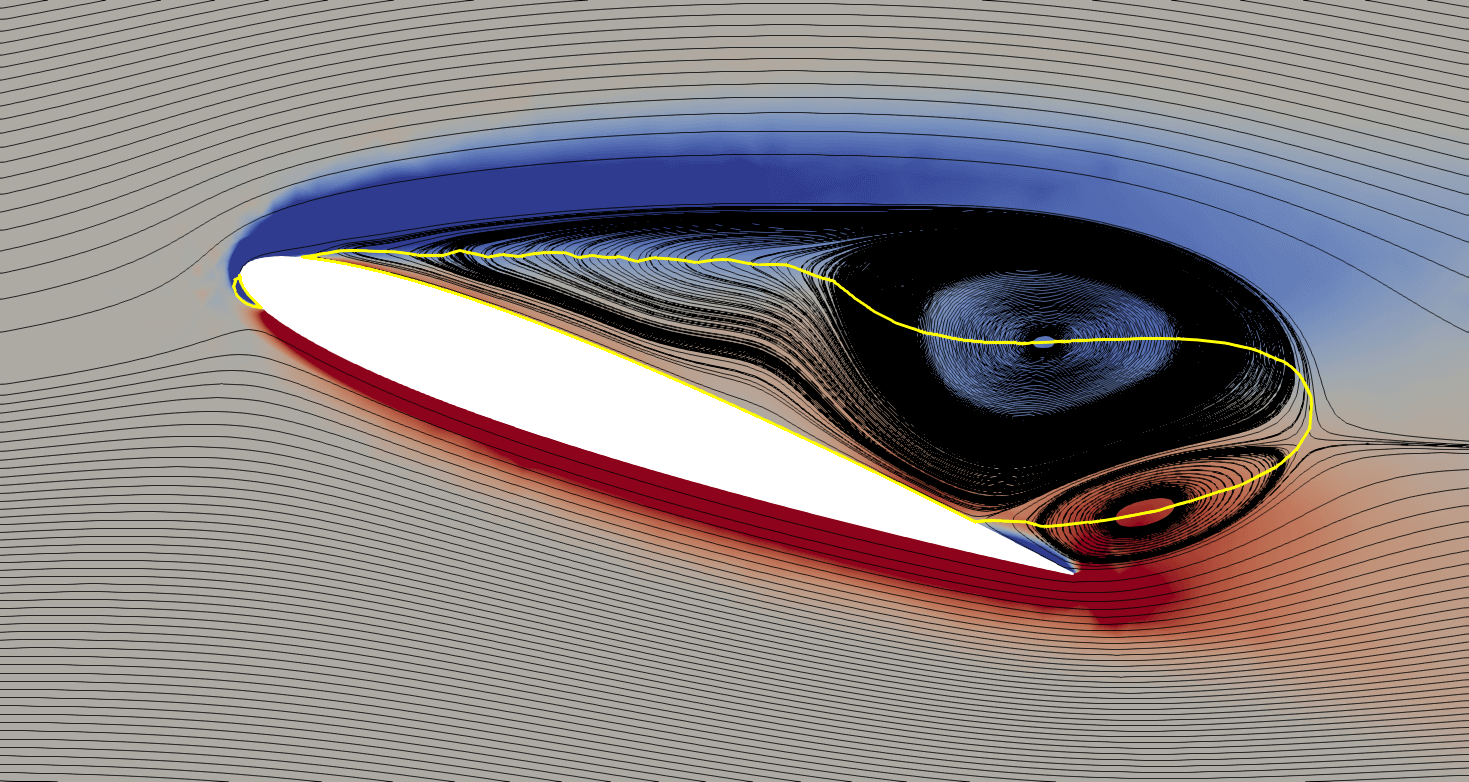}
\caption{AoA $20\degree$}
 \end{subfigure}
~
 \begin{subfigure}[b]{0.48\textwidth}
\includegraphics[width=\textwidth]{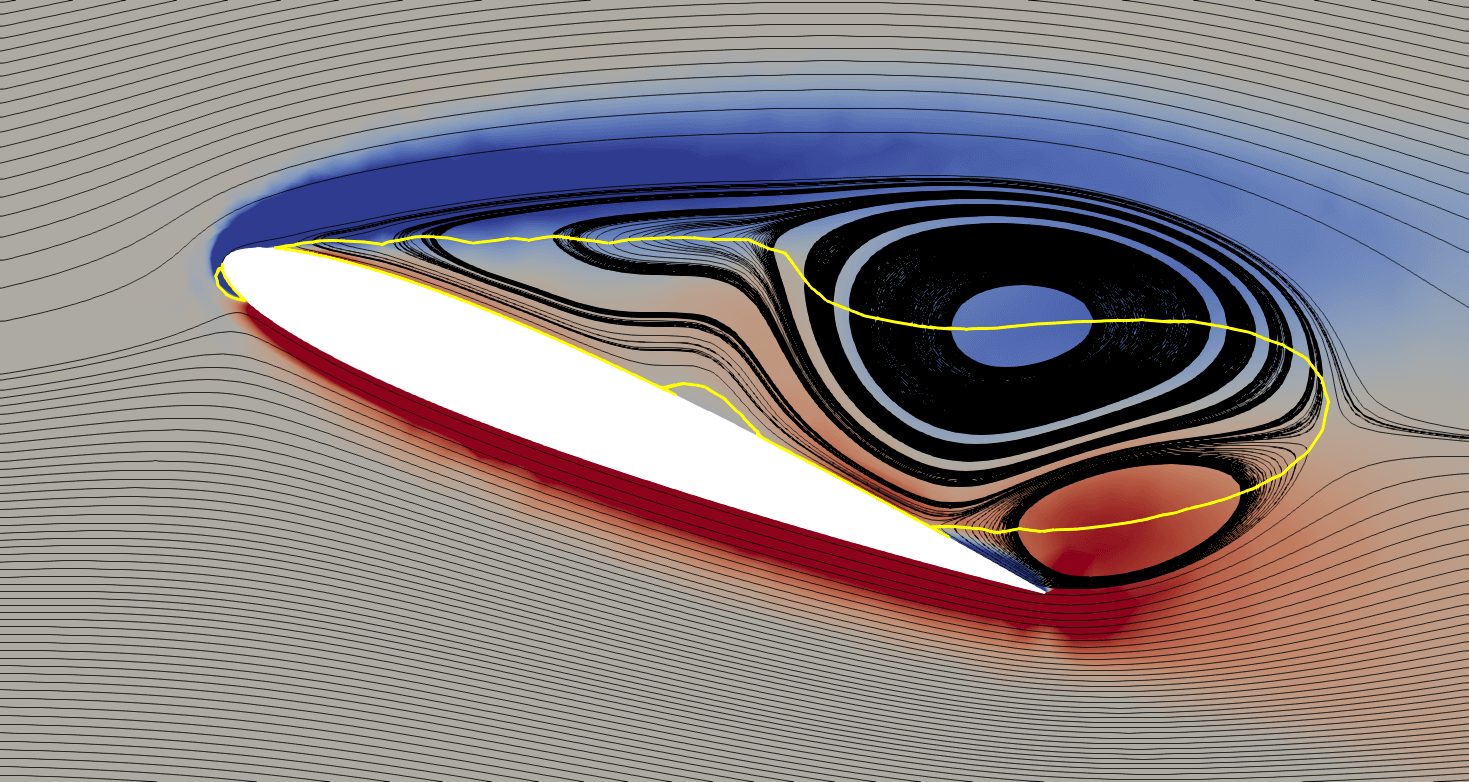}
\caption{AoA $22\degree$}
 \end{subfigure}
~
 \begin{subfigure}[b]{0.48\textwidth}
\includegraphics[width=\textwidth]{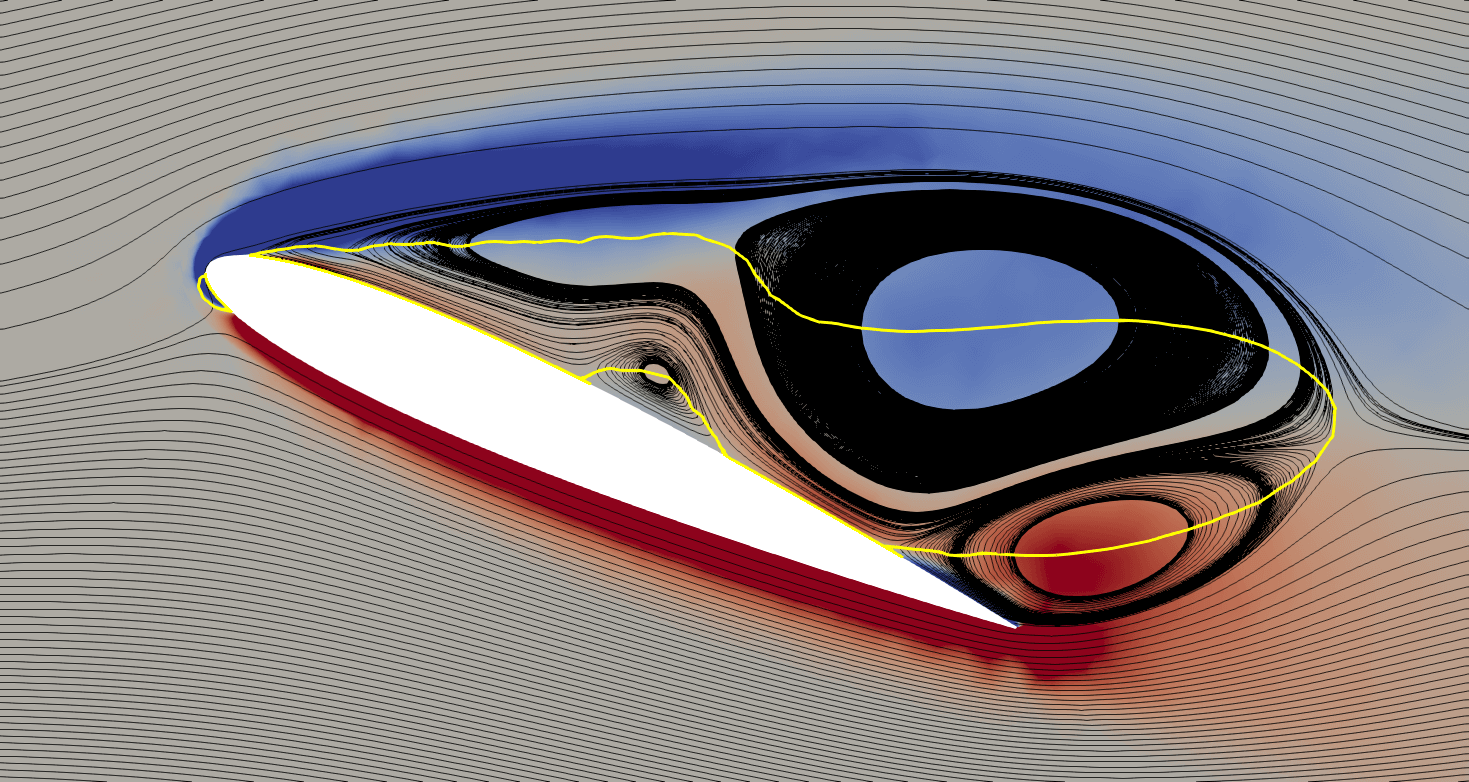}
\caption{AoA $24\degree$}
 \end{subfigure}
~
 \begin{subfigure}[b]{0.48\textwidth}
\includegraphics[width=\textwidth]{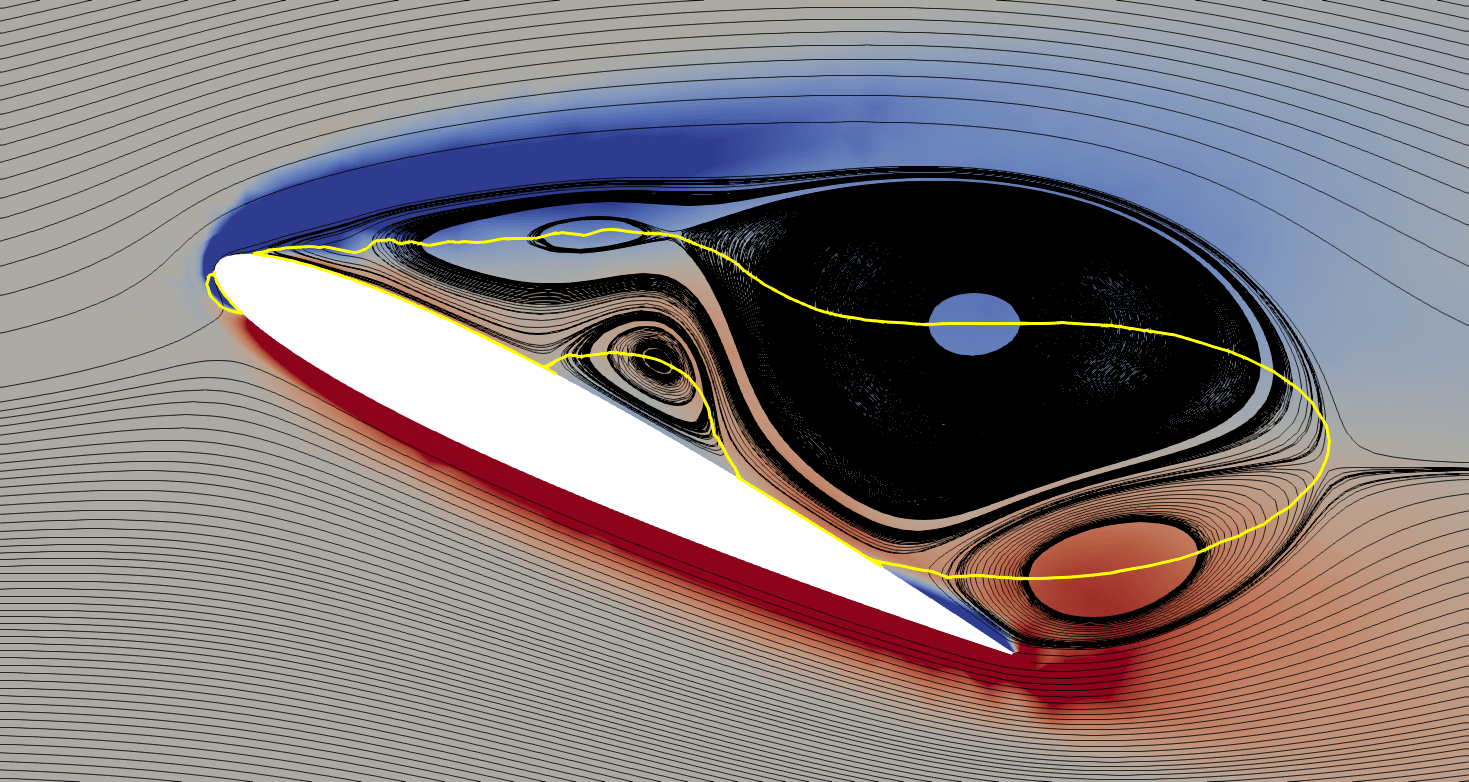}
\caption{AoA $26\degree$}
 \end{subfigure}
~
 \begin{subfigure}[b]{0.48\textwidth}
\includegraphics[width=\textwidth]{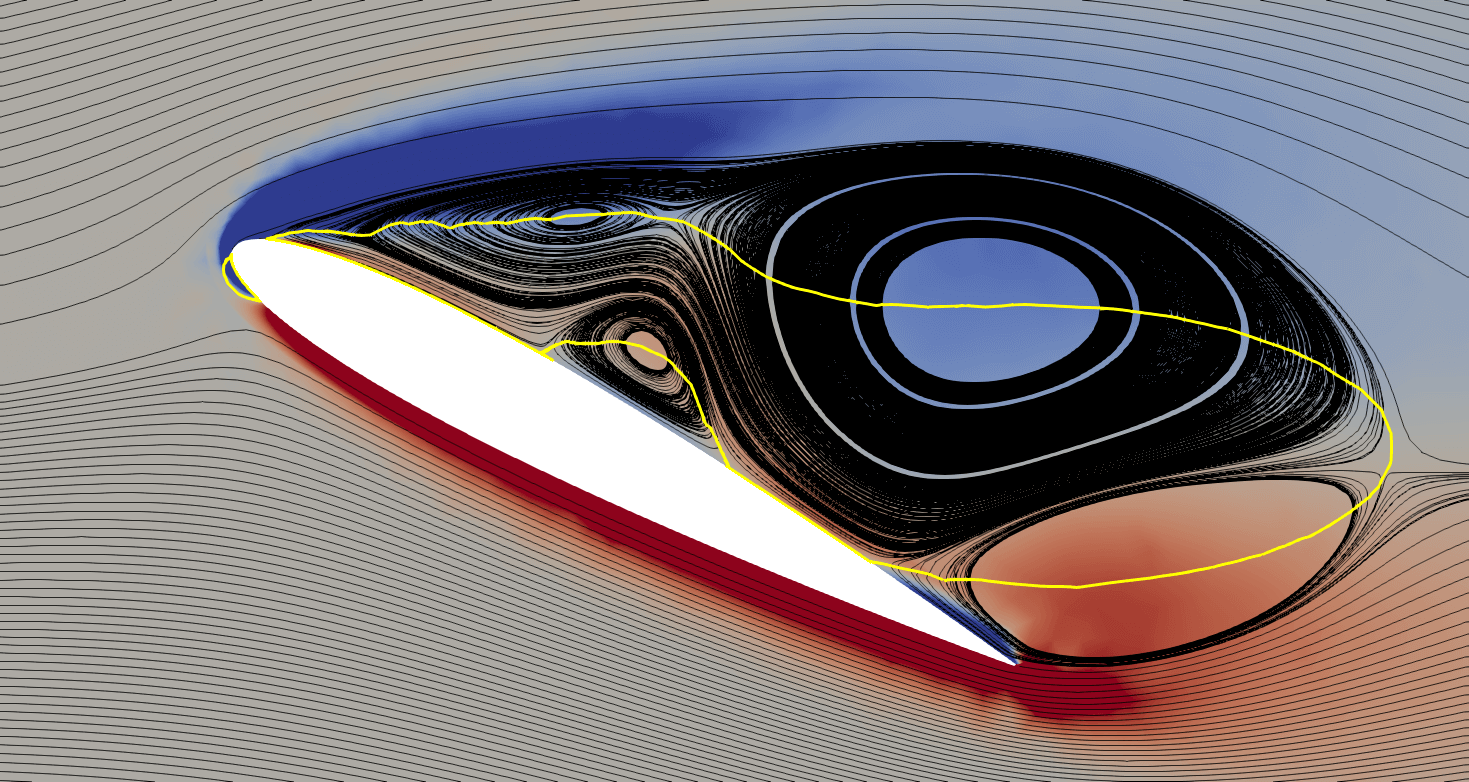}
\caption{AoA $28\degree$}
 \end{subfigure}
~
 \begin{subfigure}[b]{0.48\textwidth}
\includegraphics[width=\textwidth]{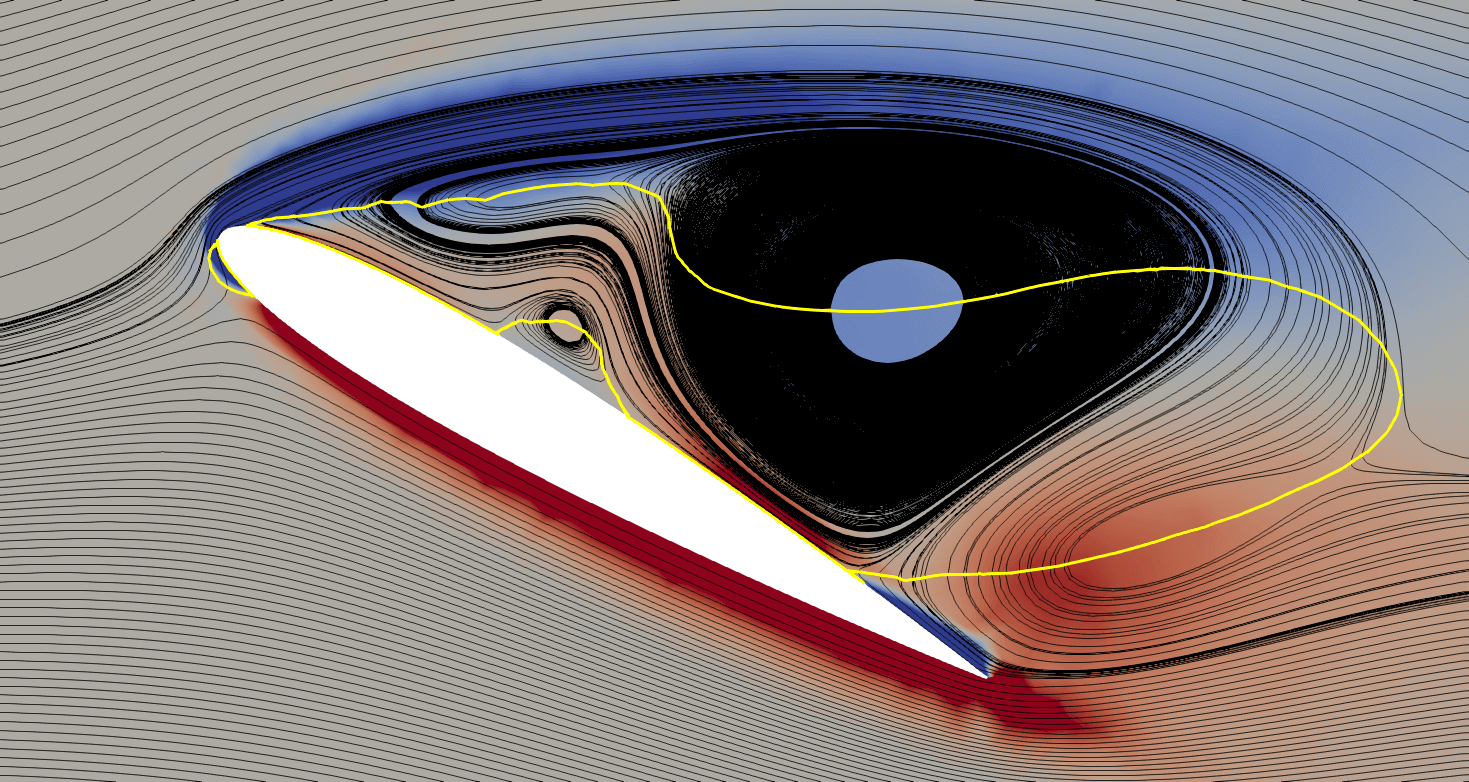}
\caption{AoA $30\degree$}
 \end{subfigure}
\caption{The time-averaged streamline and vorticity field for angles of attack $20\degree$ to $30\degree$.}\label{fig: Re1000 time-avg}
 \end{figure}

The time-averaged streamline and vorticity field are shown in Figure \ref{fig: Re1000 time-avg}. A strip of vorticity (in blue) originating from the leading edge represents the free shear layer, indicating flow separation. It is followed by the formation of the leading-edge vortex behind. In the meantime, the trailing-edge vortex (in red) is produced as the flow over the airfoil's lower surface curls upward at the trailing edge. Additionally, a secondary vortex emerges ahead of the LEV, near the upper surface of the airfoil, for angles of attack larger than $22\degree$. The appearance of this distinct vortical structure is closely related to the period-doubling behavior, which is observed at $24\degree$ (see Figure \ref{fig: phase portrait} b). From the angle of attack $26\degree$ and higher, another vortex is formed ahead of the secondary structure, referred to as the nascent LEV. The formation of this vortex is a result of a strong shear instability induced by the free shear layer and the secondary vortex. The nonlinear interplay among the three dynamical entities contributes to complex, non-periodic aerodynamic loads on the airfoil.

The yellow curves in Figure \ref{fig: Re1000 time-avg} represent the locations where the horizontal component of velocity is zero. The regions enclosed by the yellow curves correspond to flow separation or recirculation regions, and the flow within is considered reversed. It can be seen that the recirculation region expands gradually as the angle of attack increases. At an angle of attack of $30\degree$, although a secondary vortex is still observable and the nascent LEV has diminished, the earlier analysis of the phase portrait reveals a single-period orbit.

\subsection{LESP and shear layer}
To extract and quantify the leading-edge flow condition, we implement the method proposed by \citet{Deparday2022} to analyze their experimental PIV (particle image velocimetry) flow data. The dimensionless leading-edge suction parameter is given by
\begin{equation}
LESP =\frac{-\Gamma_p}{U_{\infty}~\Delta\xi}\sqrt{\frac{r_{LE}}{2c}}\left(1+\frac{1}{2}\frac{r_{LE}}{a^2} \right)^{\frac{1}{2}}. \label{eq: LESP}
\end{equation}
$\Gamma_p$ is the circulation around a circular path warping around the leading edge. $r_{LE}$ is the radius of curvature of the leading edge, and $a$ is the location of the stagnation point. To evaluate $\Gamma_p$, the circular path is chosen so that it normally intersects the leading edge surface as shown in Figure 1 of \citet{Deparday2022}. $\Delta\xi$ is the airfoil thickness between the two intersections with that circular path. Since (\ref{eq: LESP}) is derived from the inviscid potential theory, to implement it for a viscous flow data, a correction is made as
\begin{equation}
LESP =\frac{-\Gamma_p}{U_{\infty}~(\Delta\xi+\delta_{SL})}\sqrt{\frac{r_{LE}}{2c}}\left(1+\frac{1}{2}\frac{r_{LE}}{a^2} \right)^{\frac{1}{2}}, \label{eq: LESP visc}
\end{equation}
where $\delta_{SL}$ is the leading-edge shear layer thickness. This correction was made based on the conjecture that the shear layer acts as part of the airfoil shape, the flow outside the shear layer and airfoil surface is inviscid, as illustrated in Figure 4 of \citet{Deparday2022}.

\begin{figure}[h!]
\centering
 \begin{subfigure}[b]{0.4\textwidth}
\includegraphics[width=\textwidth]{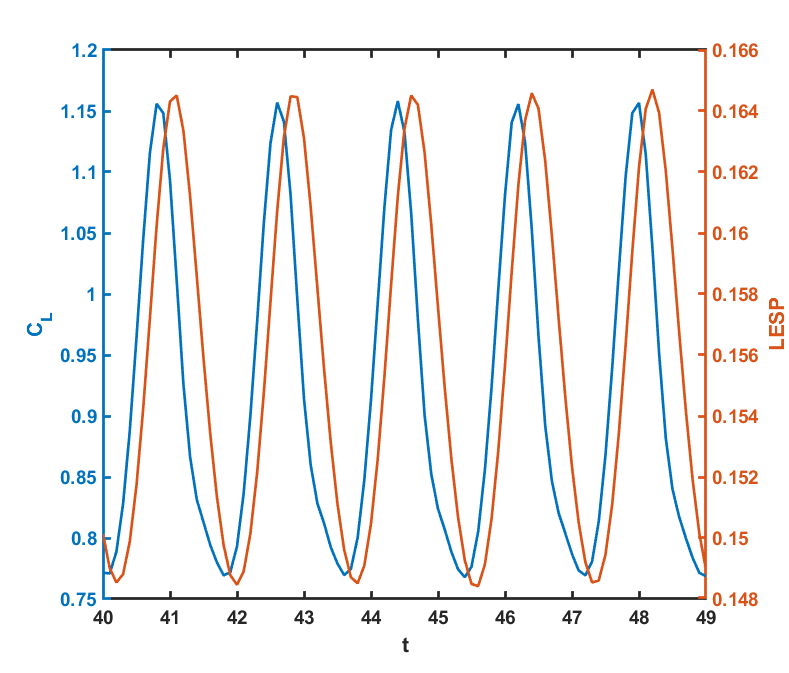}
\caption{AoA $20\degree$}
 \end{subfigure}
~
 \begin{subfigure}[b]{0.4\textwidth}
\includegraphics[width=\textwidth]{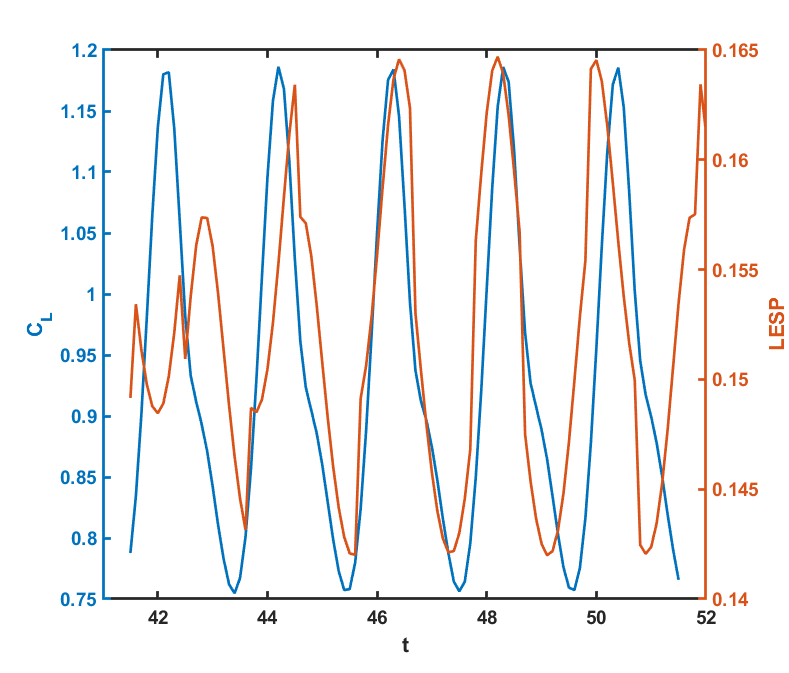}
\caption{AoA $22\degree$}
 \end{subfigure}
~
 \begin{subfigure}[b]{0.4\textwidth}
\includegraphics[width=\textwidth]{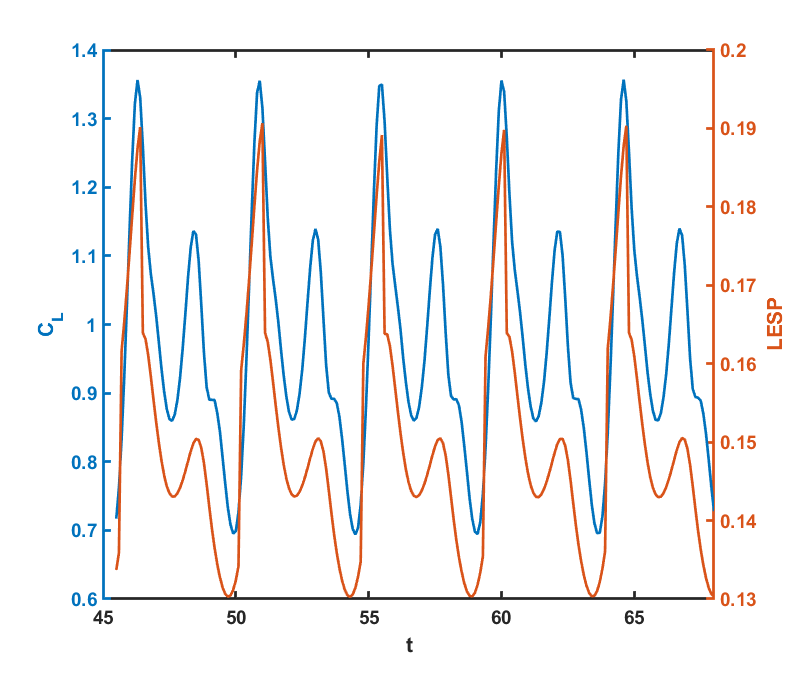}
\caption{AoA $24\degree$}
 \end{subfigure}
~
 \begin{subfigure}[b]{0.4\textwidth}
\includegraphics[width=\textwidth]{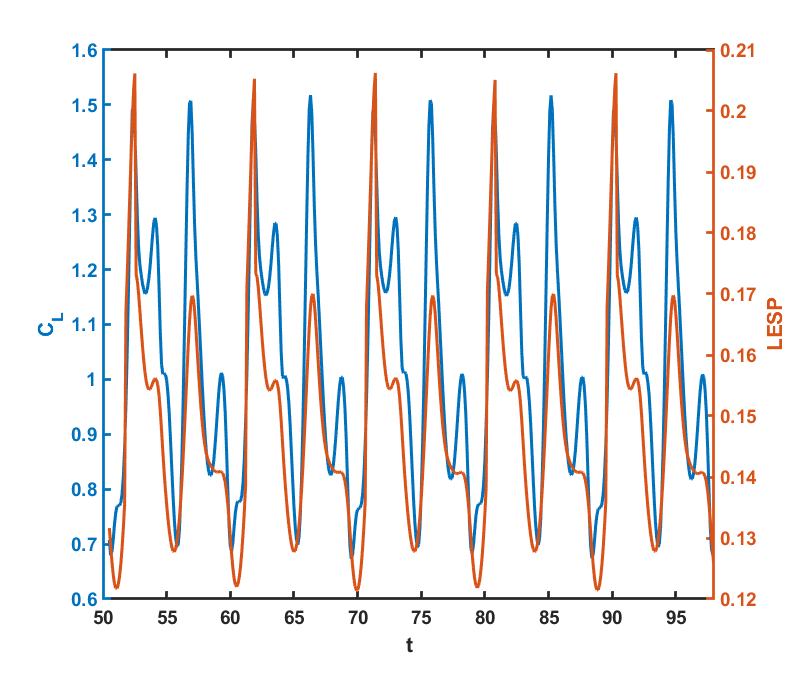}
\caption{AoA $26\degree$}
 \end{subfigure}
~
 \begin{subfigure}[b]{0.4\textwidth}
\includegraphics[width=\textwidth]{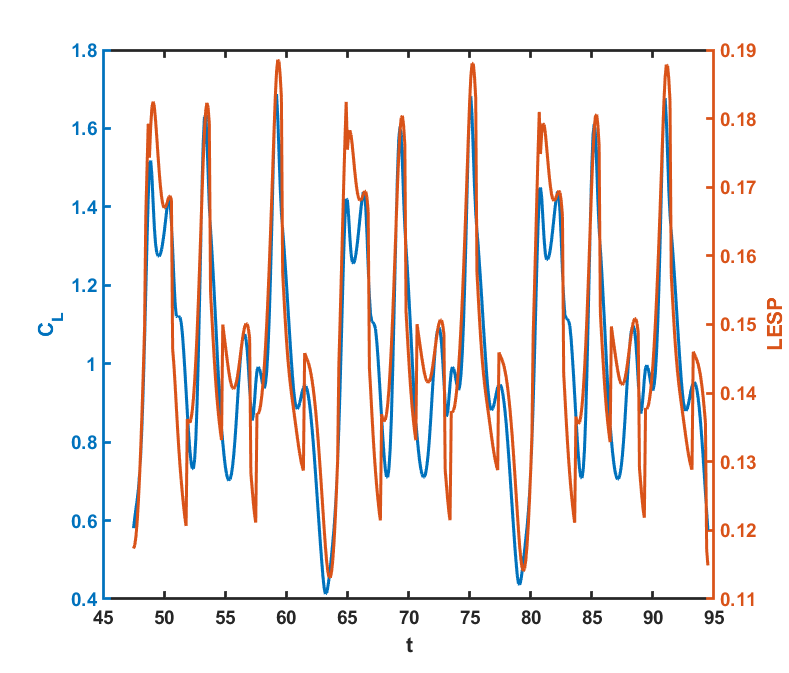}
\caption{AoA $28\degree$}
 \end{subfigure}
~
 \begin{subfigure}[b]{0.4\textwidth}
\includegraphics[width=\textwidth]{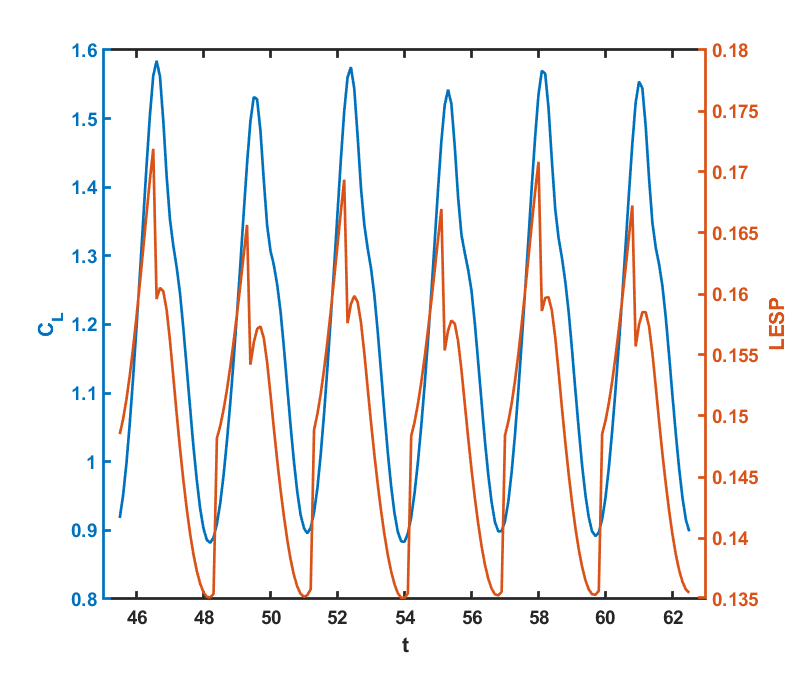}
\caption{AoA $30\degree$}
 \end{subfigure}
\caption{The instantaneous lift coefficient $C_L$ (blue) and LESP (red) for angles of attack $20\degree \sim 30\degree$.}\label{fig: Re1000 inst LESP}
 \end{figure}

Using (\ref{eq: LESP visc}), we extract the leading-edge suction parameter from our simulations; the instantaneous LESP and lift coefficient are shown in Figure \ref{fig: Re1000 inst LESP}. The data shown here exclude the startup effect of the simulation, as the simulation reached a relatively stabilized state. The instantaneous $C_L$ and LESP both exhibit periodicity in time at a angle of attack $20\degree$, the periods are almost identical while the LESP has a slight lag to the lift coefficient. In Figure \ref{fig: Re1000 inst LESP} (b), the LESP exhibits some irregularity for $t<45$ but recovers periodicity for $t$ between $45$ and $50$ before unsynchronizing with the lift for $T>50$. This might be the onset of the transition from periodic dynamics into the period-doubling regime we have discussed earlier. Both the $C_L$ and LESP contain a large and a small period for the angle of attack $24\degree$. The dynamic shifts into a more chaotic regime for an angle of attack between $26$ and $28\degree$, where both lift and the LESP appear irregular and random over time. However, we found that in this chaotic phase, the instantaneous lift coefficient and the LESP are highly synchronized. We compute the correlation coefficient between the two quantities over time in Table \ref{tab: time correl}. The results show that for angles of attack of $26$ and $28\degree$, the LESP is highly correlated with the lift coefficient, whereas in periodic regimes at other angles of attack, the correlation is lower (but still above $0.6$). The lift coefficient regains its periodicity at $30\degree$, and the LESP is also more regulated, matching the cycle of the lift coefficient.

 \begin{figure}[h!]
\centering
 \begin{subfigure}[b]{0.4\textwidth}
\includegraphics[width=\textwidth]{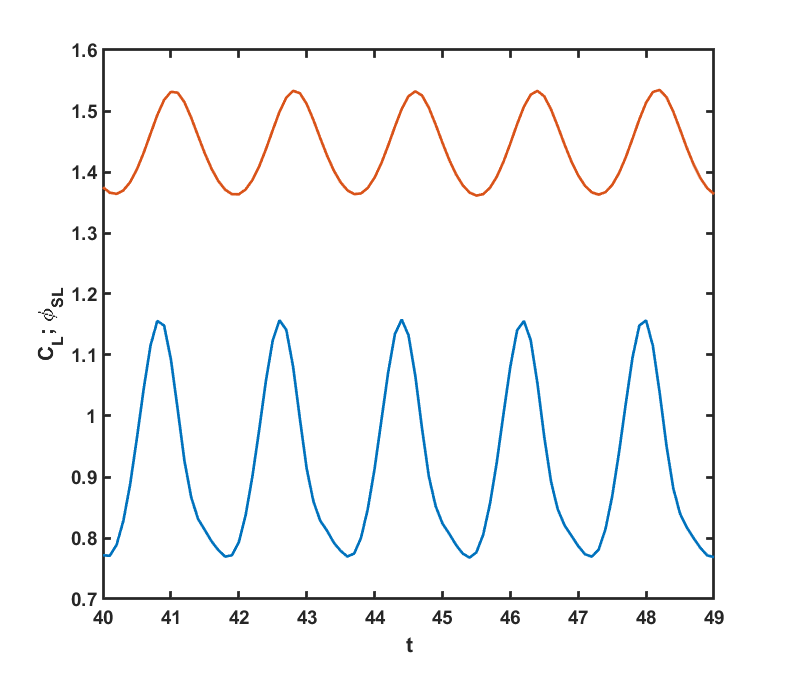}
\caption{AoA $20\degree$}
 \end{subfigure}
~
 \begin{subfigure}[b]{0.4\textwidth}
\includegraphics[width=\textwidth]{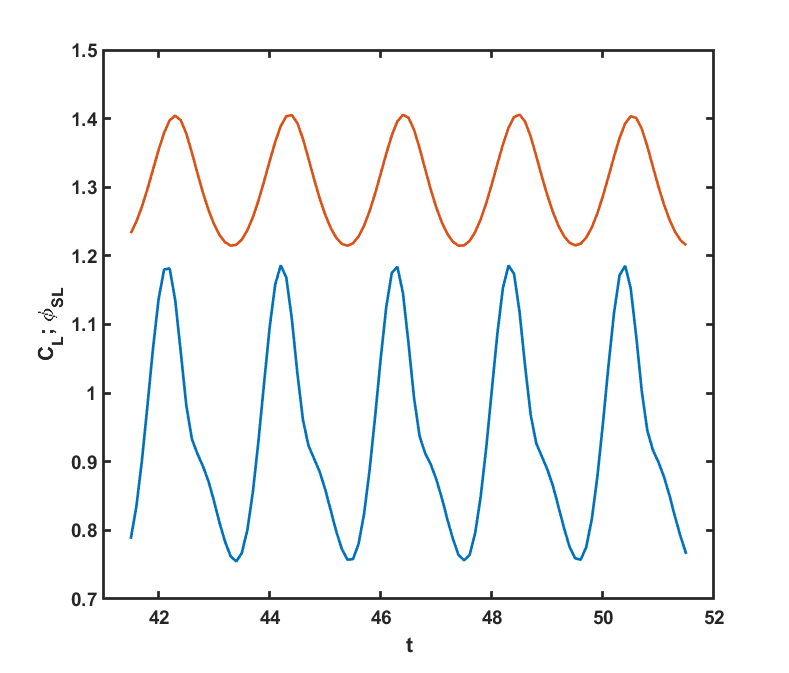}
\caption{AoA $22\degree$}
 \end{subfigure}
~
 \begin{subfigure}[b]{0.4\textwidth}
\includegraphics[width=\textwidth]{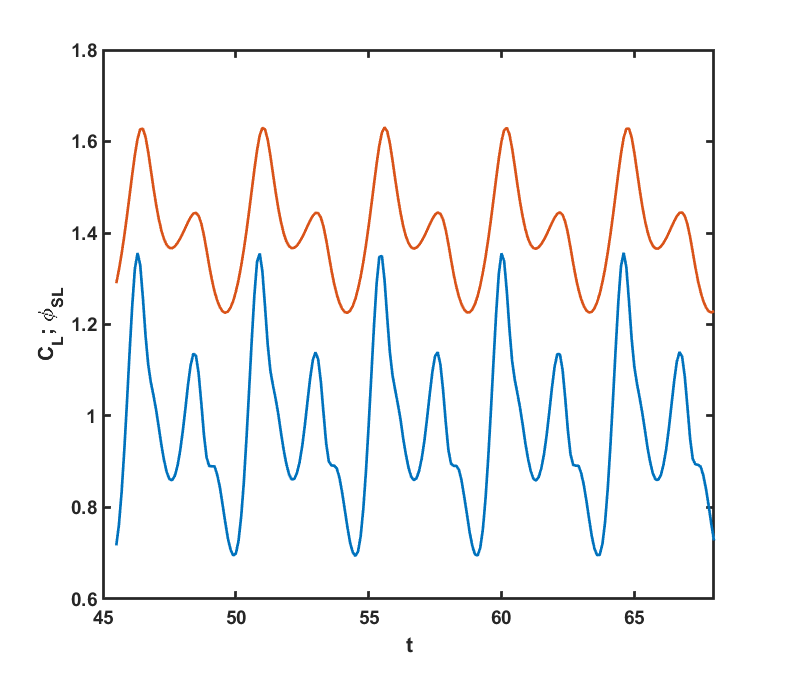}
\caption{AoA $24\degree$}
 \end{subfigure}
~
 \begin{subfigure}[b]{0.4\textwidth}
\includegraphics[width=\textwidth]{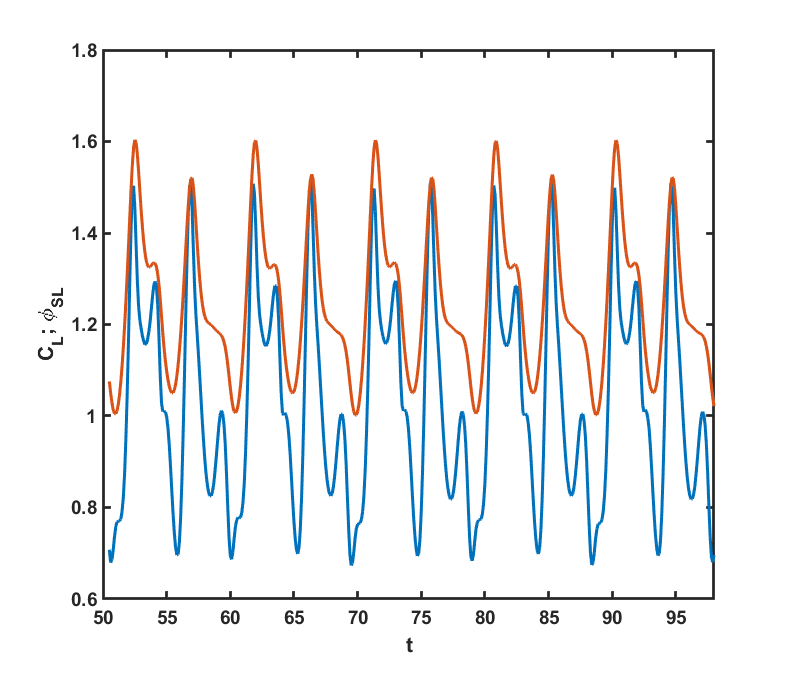}
\caption{AoA $26\degree$}
 \end{subfigure}
~
 \begin{subfigure}[b]{0.4\textwidth}
\includegraphics[width=\textwidth]{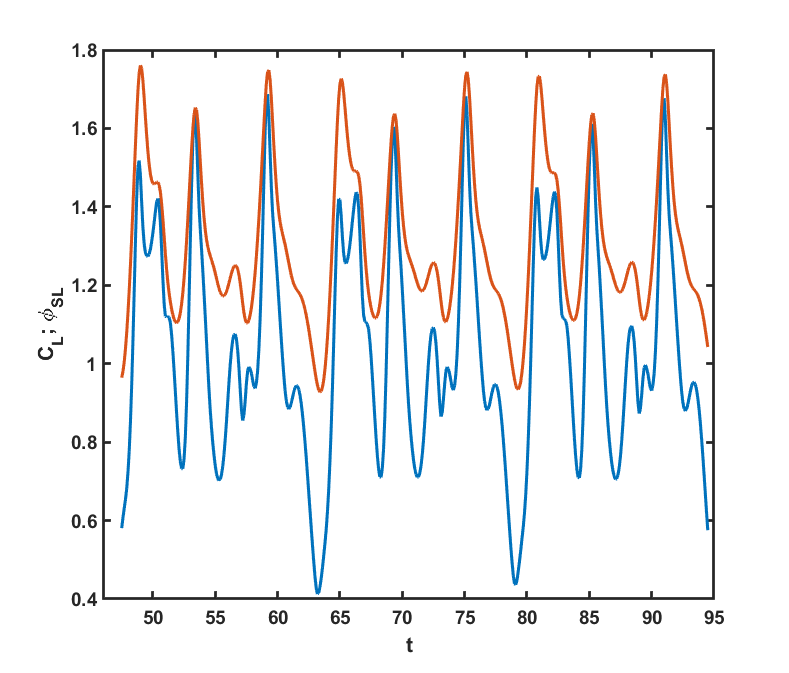}
\caption{AoA $28\degree$}
 \end{subfigure}
~
 \begin{subfigure}[b]{0.4\textwidth}
\includegraphics[width=\textwidth]{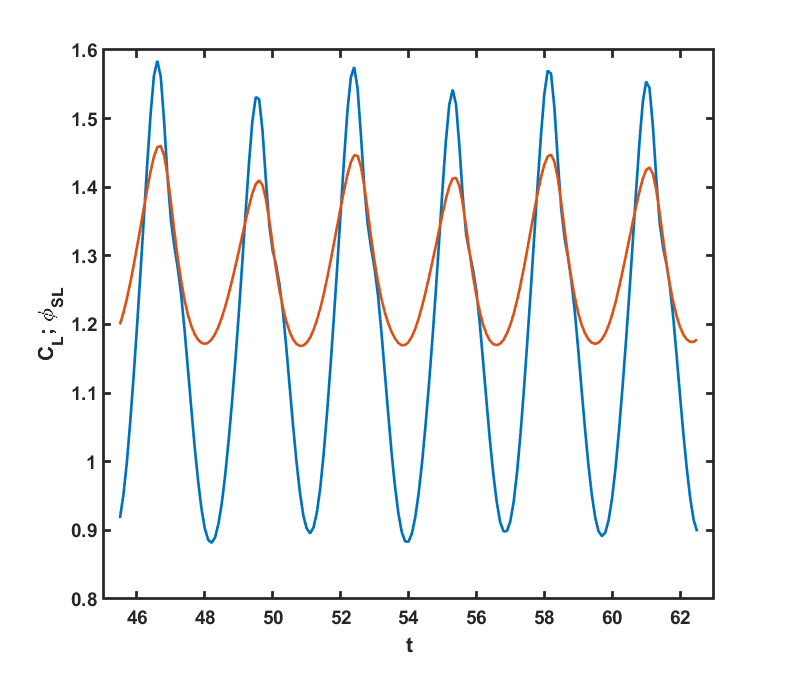}
\caption{AoA $30\degree$}
 \end{subfigure}
\caption{The instantaneous lift coefficient $C_L$ (blue) and the shear layer vorticity flux (red) for angles of attack $20\degree \sim 30\degree$.}\label{fig: Re1000 inst flux}
 \end{figure}

It is known that the vorticity contained in the LEV was fed by the shear layer, which develops and separates from the leading edge. The vorticity flux fed by the shear layer is computed by
\begin{equation}
\int_{\textit{l}} ~u \cdot \omega_z \dd y, \label{eq: flux integral}
\end{equation}
$\textit{ l}$ is a vertical line that goes from the leading edge surface and extends into the unperturbed freestream. $u$ is the $x$ component of velocity that is perpendicular to $\textit{ l}$. $\omega_z$ is the only nonzero vorticity component for a two-dimensional flow. The instantaneous vorticity flux is shown in Figure \ref{fig: Re1000 inst flux} for a couple of cycles. The vorticity flux is highly synchronized with the variation of lift in time, as evident in Table \ref{tab: time correl}, the correlation between the lift coefficient and the vorticity flux of shear layer is even higher than that of the $C_L-$LESP correlation. This can be explained by that lift is generated by the buildup of LEV \cite{PittFord2013}, which is a direct result of the injection of vorticity from the free shear layer. Therefore, the shear-layer vorticity flux has a direct influence on the lift.

\begin{table}[h!]
\centering
\caption{The correlation of lift coefficient, LESP, and the shear layer vorticity flux over time.}		\label{tab: time correl}
\captionsetup{width=\textwidth} 
\begin{tabular}{ l c c}
\toprule
Angle of attack 		& $C_L-$LESP	& $C_L-$S.L. flux\\
\midrule
$20\degree$		& 0.6521		&0.7283	\\
$22\degree$		& 0.6111		&0.9166	\\
$24\degree$		 & 0.8559 		&0.9138	\\
$26\degree$		 & 0.8793 		&0.9197	\\
$28\degree$		 & 0.8614 		&0.9190	\\
$30\degree$		 & 0.7946 		&0.9497	\\
\botrule
\end{tabular}
\end{table}

We then proceed with the time-averaged analysis where the time-averaged quantities are calculated at a fixed angle of attack. The averages are taken over five cycles and shown in Figure \ref{fig: Re1000 correl} for angles of attack $18\degree$ to $30\degree$. The time-averaged lift increases as the angle of attack increases, while the time-averaged LESP peaks at $20\degree$, then drops down to a minimum at $28\degree$ but recovers a little bit at $20\degree$. The vorticity flux of the shear layer maximizes at an angle of attack of $18\degree$, then fluctuates until $30\degree$ with an overall downward trend. We compute the correlation coefficients for the time-averaged quantities in Figure \ref{fig: Re1000 correl} over the range of angles of attack, and the result reveals that they are not well correlated. This implies that we cannot infer the time-averaged aerodynamic performance from one angle of attack to another solely from the information of LESP when the flow is laminar (Reynolds number $\sim O(10^3)$). 

 \begin{figure}[h!]
\centering
\includegraphics[width=0.6\textwidth]{CL_LESP_AoA18-30.png}
\caption{Laminar ($Re=10^3$) time-averaged $C_L$ (solid blue), LESP (solid red) and shear layer flux (dashed blue) versus angles of attack.}\label{fig: Re1000 correl}
 \end{figure}

%\subsection{Shear layer and LESP} 	\label{sec3.3}

\subsection{Turbulent regime, $Re=10^5$}
We carry out time-averaged analysis for turbulent flows. The time-averaged drag and lift coefficients have been shown in Figure \ref{fig: Re1e5 valid}. The stall appears at an angle of attack $10\degree$ while NACA0012 stalls at approximately $15\sim 16\degree$ with a typical Reynolds number of $O(10^6)$. The aerodynamic performance of the NACA0012 has been well characterized \cite{Ladson1987, Ladson1988} for Reynolds numbers ranging from $3\times10^6$ to $10^7$. Here, we set a lower Reynolds number to $100,000$ with angles of attack ranging from $0\degree$ to $30\degree$. The characteristic between the $20$ to $30\degree$ range has not been reported before, and we found that the post-stall lift coefficient steadily increases and exceeds the pre-stall lift at an angle of attack $30\degree$.

\begin{figure}[h!]
\centering
 \begin{subfigure}[b]{0.4\textwidth}
\includegraphics[width=\textwidth]{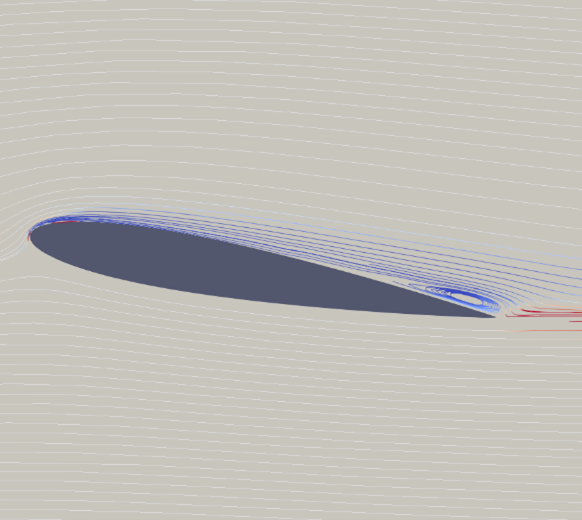}
\caption{AoA $10\degree$}
 \end{subfigure}
~
 \begin{subfigure}[b]{0.4\textwidth}
\includegraphics[width=\textwidth]{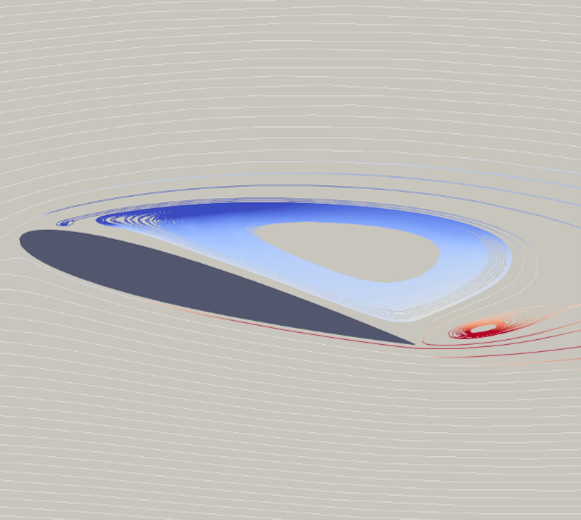}
\caption{AoA $15\degree$}
 \end{subfigure}
~
 \begin{subfigure}[b]{0.4\textwidth}
\includegraphics[width=\textwidth]{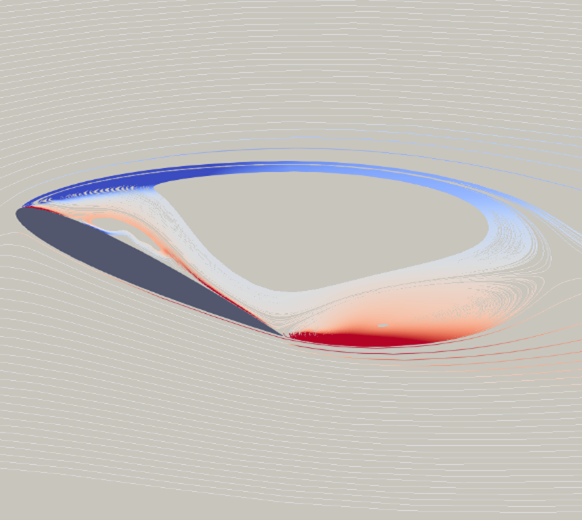}
\caption{AoA $25\degree$}
 \end{subfigure}
~
 \begin{subfigure}[b]{0.4\textwidth}
\includegraphics[width=\textwidth]{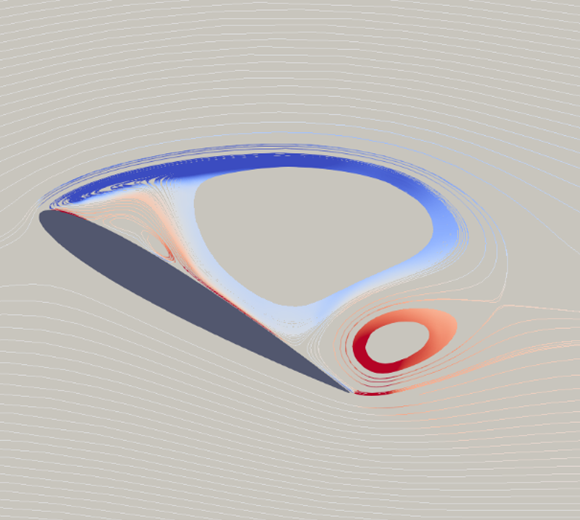}
\caption{AoA $30\degree$}
 \end{subfigure}
\caption{The time-averaged streamline and vorticity field for angles of attack.}\label{fig: Re1e5 time-avg}
 \end{figure}

\begin{figure}[h!]
\centering
\includegraphics[width=0.6\textwidth]{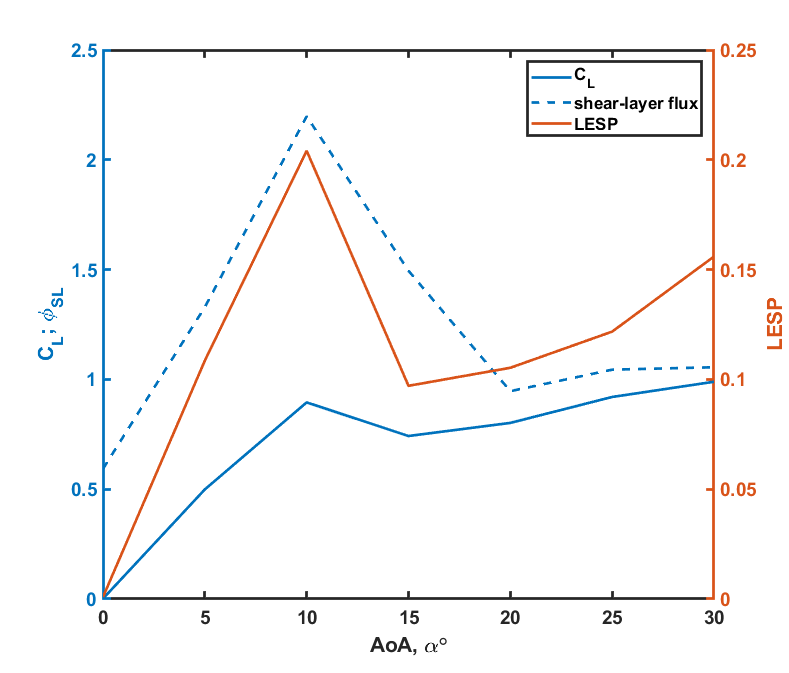}
\caption{Turbulent ($Re=10^5$) time-averaged $C_L$ (solid blue), LESP (solid red), and shear layer flux (dashed blue) versus angles of attack.}\label{fig: Re1e5 correl}
 \end{figure}

The time-averaged streamline and vorticity field are shown in Figure \ref{fig: Re1e5 time-avg}, starting from $10\degree$ angle of attack to $30\degree$. We can find that at $10\degree$ there is a small separation bubble near the trailing edge, and a massive LEV emerges at $15\degree$, indicating a separated flow and stall. The LEV expands further, and a secondary vortex appears near the airfoil surface when the angle of attack increases to $25\degree$. The secondary vortex and the trailing-edge vortex further develop while the LEV is being squeezed, and the lift gradually recovers. The time-averaged lift coefficient, LESP, and the shear-layer vorticity flux are computed in Figure \ref{fig: Re1e5 correl}. The three time-averaged quantities peak at an angle of attack of $10\degree$, then follow a similar trend, falling off to a local minimum beyond $10\degree$. The lift coefficient gradually recovers to its pre-stall value, while the LESP also climbs after the stall. The shear-layer flux also drops after the stall, but just levels. In contrast to the laminar case ($Re=1000$), we calculated the correlation coefficient over angles of attack in Table \ref{tab: AoA correl}. This implies that the time-averaged LESP may be a good indicator of the averaged lift when the angle of attack changes, particularly in a turbulence regime.

\begin{table}[h]
\caption{The correlation of lift coefficient, LESP, and the shear layer vorticity flux over angles of attack.}	\label{tab: AoA correl}
\captionsetup{width=\textwidth} 
\begin{tabular}{ l cc}
\toprule%
		 		& $C_L-$LESP	& $C_L-$S.L. flux\\
\midrule
$Re=10^3$ (laminar)		&  0.026 		& 0.4079 	\\
$Re=10^5$ (turbulent)		& 0.8576 		& 0.4790	\\
\botrule
\end{tabular}
\end{table}

\section{Conclusion}\label{sec4}
We have numerically investigated the separated flow over a NACA0012 airfoil for Reynolds numbers $1000$ and $100,000$, representing laminar and turbulent flows respectively. For Reynolds number $1000$, we analyze the time-dependent characteristics for angles of attack $20\degree$ to $30\degree$. We conclude that, within this range of angls of attack, the laminar flow field exhibits different dynamic phases, including single-period, period-doubling, and chaotic (multi-period) \cite{Chen2025}. At an angle of attack $30\degree$, it reverts to single-period mode. The two main vortices in LEV and TEV are the key structures governing lift, but their behavior is influenced by other small-scale vortices, such as secondary vortices, which affect their mutual growth and separation.

We also apply quantitative methods to CFD simulations, quantifying LESP \cite{Deparday2022} and vorticity flux to further investigate the influence of the leading-edge flow condition on the lift generated by separation. In the laminar regime $Re=1000$, the time correlation between shear-layer vorticity flux and lift coefficient, and the time correlation between the LESP and the lift both maximize around $26\degree \sim 28\degree$, where the dynamic shifts into the multi-period phase. This finding suggests that the leading-edge suction and the shear layer are linked to the lift on the airfoil when the separated flow exhibits multiple time scales. In practice, the pressure sensing technique is more widely implemented than the shear stress sensing. Therefore, monitoring the instantaneous LESP can serve as a robust indicator of the instantaneous lift. As for the time-averaged characteristic, the correlation of the averaged LESP and the averaged lift is weak among different angles of attack. The time-averaged shear-layer vorticity flux exhibits a higher correlation coefficient with lift, but it remains below 0.5. For a Reynolds number of $100,000$, the time-averaged LESP exhibits a high correlation with the time-averaged lift when varying the angle of attack. 

The present study provides a further understanding of the separated flow over an airfoil, particularly in a relatively low Reynolds number regime. Our characterization of the leading-edge suction parameter, in both instantaneous and time-averaged approaches, reveals a strong connection between the leading-edge condition and lift when the flow separates. This finding deepens our understanding of separated flow and post-stall dynamics. It may pave the way for improving the current discrete vortex method, and it also has the potential for applications in separation sensing and control.

%%%%%

\bmhead{Acknowledgements}
We thank the National Science and Technology Council, Taiwan, for the support under Grant number NSTC112-2222-E007003-MY3.

\section*{Declarations}
\begin{itemize}
\item Funding: The research leading to these results received funding from the National Science and Technology Council, Taiwan, under Grant Agreement No. NSTC112-2222-E007003-MY3
\item Competing interests: The authors have no relevant financial or non-financial interests to disclose
\item Ethics approval and consent to participate: Not applicable
\item Consent for publication: Not applicable
\item Data availability: Not applicable
\item Materials availability: Not applicable
\item Code availability: Not applicable
\item Author contribution: Ching Chang contributed to the study conception and design. Material preparation, data collection and analysis were performed by You-Peng Shih and Tang-An Li. The first draft of the manuscript was written by Ching Chang.
\end{itemize}

\noindent
If any of the sections are not relevant to your manuscript, please include the heading and write `Not applicable' for that section.

\begin{appendices}

%% clutter, break up or be distracting to the text. Appendices can consist of sections, Figures, 

\end{appendices}


\bibliography{LESP-ref}% common bib file

\end{document}